\documentclass[aps, pra, twocolumn, amsmath, amssymb, superscriptaddress, nofootinbib]{revtex4-1}

\makeatletter
\def\p@subsection{}
\def\p@subsubsection{}
\makeatother

\usepackage[utf8]{inputenc}
\usepackage{graphicx}
\usepackage{units}
\usepackage{color}
\usepackage[pdftex, colorlinks=true, linkcolor=myblue, citecolor=myblue,
urlcolor=myblue]{hyperref}
\usepackage{times}
\usepackage{comment}
\usepackage{graphicx}
\usepackage{amsmath, calc, cases}
\usepackage{mathrsfs}
\usepackage{amsfonts}
\usepackage{amssymb}
\usepackage{bm}
\usepackage{bbm}
\usepackage{color}
\usepackage{array}
\usepackage{units}

\usepackage{amsmath,empheq}

\definecolor{myblue}{rgb}{0,0,1}
\definecolor{myred}{rgb}{1,0,0}

\DeclareMathOperator{\Tr}{Tr}

\DeclareMathOperator\arctanh{arctanh}

\begin{document}


\title{Energetics of a pulsed quantum battery}


\author{Charles Andrew Downing} 
\email{c.a.downing@exeter.ac.uk}
\affiliation{Department of Physics and Astronomy, University of Exeter, Exeter EX4 4QL, United Kingdom}

\author{Muhammad Shoufie Ukhtary}
\affiliation{Department of Physics and Astronomy, University of Exeter, Exeter EX4 4QL, United Kingdom}
\affiliation{Research Center for Quantum Physics, National Research and Innovation Agency (BRIN), South Tangerang 15314, Indonesia}


\date{\today}


\begin{abstract}
\noindent \textbf{Abstract}\\
The challenge of storing energy efficiently and sustainably is highly prominent within modern scientific investigations. Due to the ongoing trend of miniaturization, the design of expressly quantum storage devices is itself a crucial task within current quantum technological research. Here we provide a transparent analytic model of a two-component quantum battery, composed of a charger and an energy holder, which is driven by a short laser pulse. We provide simple expressions for the energy stored in the battery, the maximum amount of work which can be extracted, both the instantaneous and the average powers, and the relevant charging times. This allows us to discuss explicitly the optimal design of the battery in terms of the driving strength of the pulse, the coupling between the charger and the holder, and the inevitable energy loss into the environment. We anticipate that our theory can act as a helpful guide for the nascent experimental work building and characterizing the first generation of truly quantum batteries.
\end{abstract}


\maketitle



\noindent \textbf{Introduction.} Battery research seeks to play a role in addressing some of the most pressing issues affecting modern society, including the sustainable generation, storage and transport of energy. These kinds of important problems also arise in the quantum world, where the very nature of fields like quantum information and quantum thermodynamics suggest the existence of a novel type of energy storage device: a quantum battery~\cite{Alicki2013, Campaioli2023}. 

Early experimental work (mostly with spins and superconducting qubits) has already started to pioneer the fundamental workings of quantum batteries, including: the degradation of the stored energy over time, various charging and discharging protocols, and the overall power performance~\cite{Peterson2019, Cimini2020, Quach2022, Joshi2022, Hu2022, Stevens2022, Maillette2022}. Meanwhile, theoretical studies have sought to provide innovative ideas to increase the efficiency and precision of quantum batteries, as well as to exploit various inherently quantum advantages~\cite{Ferraro2018, Andolina2018, Farina2019, Crescente2020, NewCrescente2020, Santos2020, Carrega2020, Santos2021, Gyhm2022, Shaghaghi2002, Catalano2023}.

Here we seek to provide a simple yet explanatorily powerful model of a bipartite quantum battery driven by a short laser pulse. We model our battery charger and battery holder as quantum harmonic oscillators, such that our theory contributes to the emerging body of work on quantum continuous variable batteries~\cite{Friis2018, Huangfu2021, Downing2023, Centrone2021, Konar2022}. Within an open quantum systems approach, we provide compact expressions describing the energetic and power performance of the battery, as well as brief formulae for the maximum amount of work which can be extracted~\cite{Alicki2013, Allahverdyan2004} and for various desirable charging times. Interestingly, the behaviour of the pulsed quantum battery is intrinsically governed by the presence of an exceptional point~\cite{Berry2003}, which are becoming increasingly influential in both classical and quantum optics~\cite{Miri2019, Downing2021, Bender2023}. We note that theoretical analyses of other driving protocols can be found in Refs.~\cite{Crescente2020, NewCrescente2020} amongst others.
\\

\begin{figure}[tb]
 \includegraphics[width=0.8\linewidth]{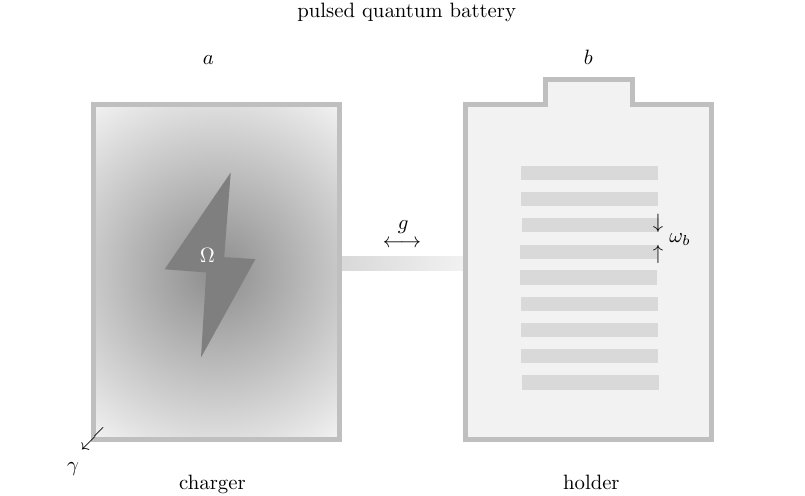}
 \caption{ \textbf{Sketch of the pulsed quantum battery.} The bipartite quantum battery is composed of a battery charger (associated with the bosonic operator $a$) and a battery holder (operator $b$). Both parts are modelled as quantum harmonic oscillators with the level spacing $\omega_b$, and they are coupled at the strength $g$. The charger is driven by a pulse of strength $\Omega$, and it suffers from dissipation at the rate $\gamma$.}
 \label{sketchy}
\end{figure}


\noindent \textbf{Model.} The total Hamiltonian operator $\hat{H}$ describing the bipartite quantum battery can be split into four components [cf. the cartoon of Fig.~\ref{sketchy}]
\begin{equation}
\label{eq:Haxcsdfdsfdsfsdfvcxvmy}
 \hat{H} =  \hat{H}_a +  \hat{H}_b +  \hat{H}_{a-b} +  \hat{H}_{d}.
\end{equation}
The battery charger and battery holder are taken to be quantum harmonic oscillators, governed by the twin terms (we set $\hbar = 1$ here and throughout)
\begin{equation}
\label{eq:assdd}
\hat{H}_a = \omega_b a^\dagger a,
\quad\quad\quad
\hat{H}_b = \omega_b b^\dagger b,
\end{equation}
where $\omega_b$ is the energy level spacing of each oscillator. The operator $a^\dagger$ creates an excitation in the battery charger and $a$ destroys an excitation, while $b^\dagger$ and $b$ do likewise in the battery holder. The bosonic commutation relations $[a, a^\dagger] = 1$ and $[b, b^\dagger] = 1$ are observed. The interaction between the charger and the holder is modelled with the coupling term
\begin{equation}
\label{eq:assdsdd}
\hat{H}_{a-b} = g \left( a^\dagger b + b^\dagger a \right),
\end{equation}
where the coupling frequency is of strength $g \ge 0$. The battery charger is driven at the time $t = 0$ by a short laser pulse
\begin{equation}
\label{eq:assd}
\hat{H}_d = \Omega ~ \delta ( t ) \left(  a^\dagger + a \right),
\end{equation}
which is of dimensionless strength $\Omega \ge 0$ and where $\delta (t)$ is Dirac's delta function (we support this choice of drive in the supplementary data). We assume that the battery holder is sufficiently isolated from the environment such that its energy losses are negligible, as should be the case for any meaningful energy storage device. However, this isolation implies that the battery holder is rather hard to transfer energy into in the first place. Therefore we couple the battery holder to the battery charger, which is easy to drive and hence it also feels its external environment such that it noticeably dissipates [cf. the sketch of Fig.~\ref{sketchy}]. These losses from the battery charger are incorporated into our model using a quantum master equation, which is considered to be in Gorini–Kossakowski–Sudarshan–Lindblad form as~\cite{Breuer2002}
\begin{equation}
\label{eq:xcxdsfdsfc3}
\partial_t \rho = \mathrm{i} \left[ \rho, \hat{H} \right] + \frac{\gamma}{2} \left( 2 a \rho a^\dagger  - a^\dagger a \rho - \rho  a^\dagger a  \right),
\end{equation}
where the quantum battery system's density matrix is $\rho$, the battery charger dissipation rate $\gamma \ge 0$, and the total Hamiltonian operator $\hat{H}$ is defined in Eq.~\eqref{eq:Haxcsdfdsfdsfsdfvcxvmy}. There are a triumvirate of measures of primary interest for the proposed pulsed quantum battery: the energy $E$ stored in the battery holder, the corresponding instantaneous power $\mathcal{P}$, and the average power $P$. These three quantities are given in terms of the average population of the battery holder $\langle b^\dagger b \rangle$, and are defined by
\begin{equation}
\label{eq:xcxc}
E = \omega_b \langle b^\dagger b \rangle,
\end{equation}
\begin{equation}
\label{eq:xcxc2}
\mathcal{P} = \frac{\mathrm{d}}{\mathrm{d}t} E,
\end{equation}
\begin{equation}
\label{eq:xcxc3}
P = \frac{E}{t}.
\end{equation}
In particular, we seek to find the optimal values of the model parameters $g$, $\Omega$ and $\gamma$ [as represented pictorially in Fig.~\ref{sketchy}] such that desired energetics (as judged by $E$, $\mathcal{P}$ and $P$) are achieved in the smallest possible elapsed time after the battery charger is driven, such that the charger can then be disconnected.
\\


\begin{figure*}[tb]
 \includegraphics[width=\linewidth]{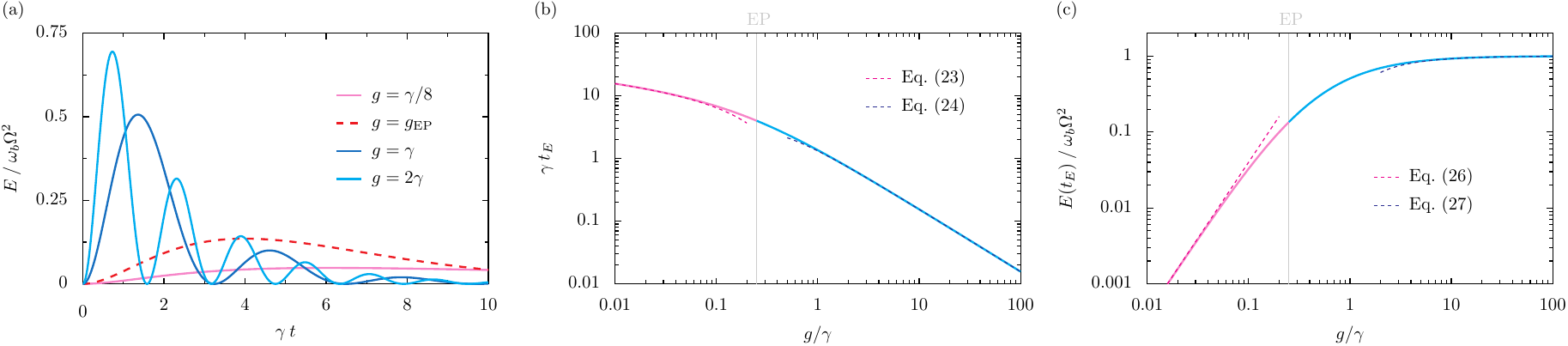}
 \caption{ \textbf{Energetics of the pulsed quantum battery.} Panel (a): the dynamical energy $E$ stored in the battery (in units of $\omega_b \Omega^2$) as a function of the time $t$ (in units of the inverse decay rate of the charger $\gamma^{-1}$) [cf. Eq.~\eqref{eq:xcvxcv}]. Coloured lines: results for increasingly large strengths of the charger-battery coupling $g$, from below the exceptional point (pink line) to exactly at the exceptional point (dashed red line) to above the exceptional point (blue and cyan lines). Panel (b): the time $t_E$, corresponding to the maximum of the stored energy $E$, as a function of the coupling-to-dissipation ratio $g/\gamma$ is represented with the pink-cyan line [cf. Eq.~\eqref{eq:xcdvzdvdvdvxcv}]. Dashed lines: asymptotic results at small and large couplings [cf. Eq.~\eqref{eq:srvgrs} and Eq.~\eqref{eq:sfdfdsvefg}]. Panel (c): the maximum energy $E(t_E)$ as a function of $g/\gamma$ is represented with the pink-cyan line [cf. Eq.~\eqref{eq:xcxdxvdvxdvxdvvxcv}]. Dashed lines: asymptotic results at small and large couplings [cf. Eq.~\eqref{eq:sfddfs} and Eq.~\eqref{eq:sdfdfsdf}]. Vertical grey lines in panels (b) and (c): guides for the eye at the exceptional point of $g = \gamma/4$ [cf. Eq.~\eqref{eq:sdfdfds}].}
 \label{Efig}
\end{figure*}

\noindent \textbf{Dynamics.} The first moments of the pulsed quantum battery follow from the trace property $\Tr ( \mathcal{O} \rho ) = \langle \mathcal{O} \rangle$, which is valid for any operator $\mathcal{O}$, and the quantum master equation of Eq.~\eqref{eq:xcxdsfdsfc3}. This averaging process leads to the equation of motion (see the supplementary data for details)
\begin{equation}
\label{eq:xcxcxxvcxdsfdsfc3}
\mathrm{i} \partial_t \psi = \mathcal{H} \psi + \mathcal{S},
\end{equation}
where the first moments contained within $\psi$ and the dynamical matrix $\mathcal{H}$ are together given by
\begin{equation}
\label{eq:sdadxcvxcvxvczdcds}
 \psi =
\begin{pmatrix}
 \langle a \rangle \\
\langle b \rangle
\end{pmatrix}, 
\quad\quad\quad
 \mathcal{H} =
\begin{pmatrix}
 \omega_b - \mathrm{i} \frac{\gamma}{2} & g   \\
 g & \omega_b  
\end{pmatrix},
\end{equation}
while the pulsed drive of the charger appears in the final term
\begin{equation}
\label{eq:sdadczdcds}
 \mathcal{S} =
 \Omega
\begin{pmatrix}
 \delta ( t )\\
0
\end{pmatrix}.
\end{equation}
The two complex eigenvalues $\epsilon_{\pm}$ of the non-Hermitian matrix $\mathcal{H}$ defined in Eq.~\eqref{eq:sdadxcvxcvxvczdcds} are
\begin{equation}
\label{eq:sdads}
\epsilon_{\pm} = \omega_b - \mathrm{i} \frac{\gamma}{4} \pm G,
\end{equation}
where the renormalized coupling rate $G$ is defined by
\begin{equation}
\label{eq:zxczxczxzx}
 G = \sqrt{ g^2 - \left( \tfrac{\gamma}{4} \right)^2 }.
\end{equation}
Notably, the pulsed quantum battery exhibits a quantum spectral degeneracy~\cite{Vidiella2023, Fox2023} at a certain exceptional point, which occurs when $G = 0$ (or equivalently when $g = g_{\mathrm{EP}}$) where
\begin{equation}
\label{eq:sdfdfds}
g_{\mathrm{EP}} = \frac{\gamma}{4}.
\end{equation}
The existence of this exceptional point divides the response of the pulsed quantum battery into two distinct behavioural regimes, a classification which is used in what follows. At all times $t >0$, that is after the driving pulse has been applied, the solution of the then Schr\"{o}dinger-like equation of motion of Eq.~\eqref{eq:xcxcxxvcxdsfdsfc3} reads
\begin{equation}
\label{eq:zcxzxczxc}
 \psi = A \begin{pmatrix}
- \mathrm{i} G -\frac{\gamma}{4}  \\
- \mathrm{i} g
\end{pmatrix} \mathrm{e}^{ - \mathrm{i} \epsilon_{+} t}
+
B \begin{pmatrix}
 \mathrm{i} G -\frac{\gamma}{4} \\
- \mathrm{i} g
\end{pmatrix} \mathrm{e}^{ - \mathrm{i} \epsilon_{-} t},
\end{equation}
where the eigenvalues $\epsilon_{\pm}$ are defined in Eq.~\eqref{eq:sdads}. The two constants $A$ and $B$ can be found by integrating Eq.~\eqref{eq:xcxcxxvcxdsfdsfc3} with respect to time over the small temporal interval from $t = -\varepsilon$ to $t = +\varepsilon$, which contains the pulse exactly at $t = 0$ [cf. Eq.~\eqref{eq:sdadczdcds}]. Taking the limit of $\varepsilon \to 0$ reveals that the values of the constants which satisfy the boundary condition are
\begin{equation}
\label{eq:zxczxczxcvxcvxzx}
A = - B,
\quad\quad\quad\quad
A = \frac{\Omega}{2 G}.
\end{equation}
Therefore, the solution of Eq.~\eqref{eq:zcxzxczxc} provides the dynamical behaviour of the mean value of the operator $a$ as follows
\begin{widetext}
\begin{equation}
\label{eq:adscdaadd}
 \langle a \rangle = \begin{cases}
  -\mathrm{i} \Omega \left[ \cosh \left( \Gamma t \right) - \frac{\gamma}{4\Gamma} \sinh \left( \Gamma t \right) \right] \mathrm{e}^{- \frac{\gamma t}{4}} \mathrm{e}^{-\mathrm{i} \omega_b t}, & g < g_{\mathrm{EP}}, \\
  -\mathrm{i} \Omega \left( 1 - \frac{\gamma t}{4} \right) \mathrm{e}^{- \frac{\gamma t}{4}} \mathrm{e}^{-\mathrm{i} \omega_b t}, & g = g_{\mathrm{EP}}, \\
-\mathrm{i} \Omega \left[ \cos \left( G t \right) - \frac{\gamma}{4G} \sin \left( G t \right) \right] \mathrm{e}^{- \frac{\gamma t}{4}} \mathrm{e}^{-\mathrm{i} \omega_b t},  & g > g_{\mathrm{EP}} ,
\end{cases}
\end{equation}
\end{widetext}
where we have made use of the exceptional point of Eq.~\eqref{eq:sdfdfds} in order to categorize the response of the system. In particular, below the exceptional point of $g_{\mathrm{EP}}$ where the behaviour is non-oscillatory, we have introduced the renormalized dissipation rate
\begin{equation}
\label{eq:opj}
  \Gamma = \sqrt{  \left( \tfrac{\gamma}{4} \right)^2 - g^2 },
\end{equation}
as the counterpart to $G$, as was defined in Eq.~\eqref{eq:zxczxczxzx}. Above the exceptional point the battery charger displays Rabi-like oscillations, while exactly at $g = g_{\mathrm{EP}}$ some remarkable power function dynamics is instead presented following Eq.~\eqref{eq:adscdaadd}. Similarly, the first moment of the battery holder can be read off from Eq.~\eqref{eq:zcxzxczxc} as
\begin{equation}
\label{eq:zcszsc}
 \langle b \rangle = \begin{cases}
-\Omega \: \frac{g}{\Gamma} \sinh \left( \Gamma t \right) \mathrm{e}^{- \frac{\gamma t}{4}} \mathrm{e}^{-\mathrm{i} \omega_b t},   & g < g_{\mathrm{EP}}, \\
-\Omega \: \frac{\gamma t}{4} ~ \mathrm{e}^{- \frac{\gamma t}{4}} \mathrm{e}^{-\mathrm{i} \omega_b t}, & g = g_{\mathrm{EP}}, \\
-\Omega \: \frac{g}{G} \sin \left( G t \right) \mathrm{e}^{- \frac{\gamma t}{4}} \mathrm{e}^{-\mathrm{i} \omega_b t},  & g > g_{\mathrm{EP}},
\end{cases}
\end{equation}
which manifests a similar structure to the solutions of Eq.~\eqref{eq:adscdaadd}, including a non-oscillatory to oscillatory transition at the exceptional point (where the response is linear in time). The solution of Eq.~\eqref{eq:zcszsc} allows for the mean population of the battery holder to be accessed thanks to the relation $\langle b^\dagger \rangle  \langle b \rangle =  \langle b^\dagger b \rangle$, which holds within this particular model because the joint charger–battery holder system evolves in a product state~\cite{Farina2019, Downing2023}. This correlator factorization prescription allows for the energetics of the quantum battery to be readily investigated.
\\


\begin{figure*}[tb]
 \includegraphics[width=\linewidth]{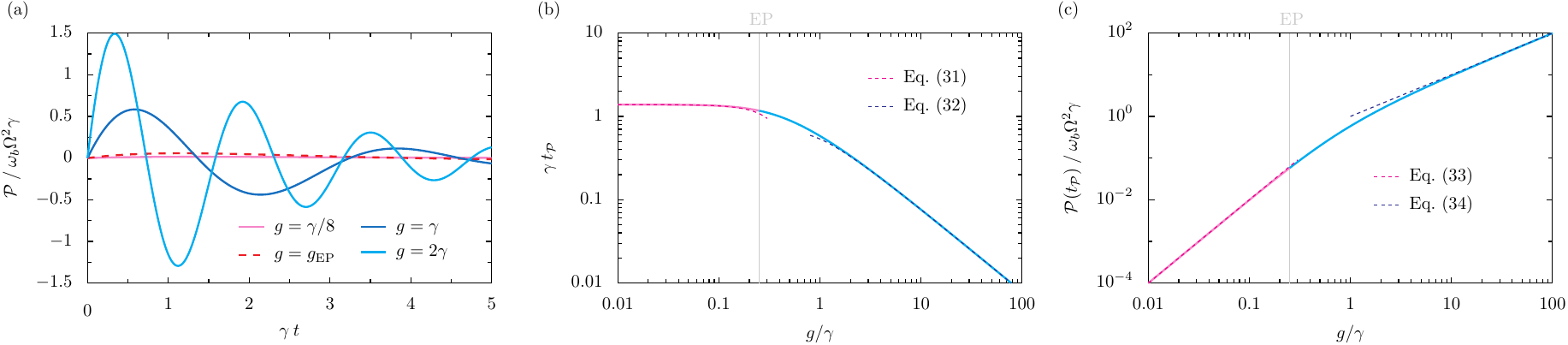}
 \caption{ \textbf{Instantaneous power of the pulsed quantum battery.} Panel (a): the instantaneous power $\mathcal{P}$ of the battery (in units of $\omega_b \Omega^2 \gamma$) as a function of the time $t$ (in units of the inverse decay rate of the charger $\gamma^{-1}$) [cf. Eq.~\eqref{eq:xcvzxczcxzxcv}]. Coloured lines: results for increasingly large strengths of the charger-battery coupling $g$, from below the exceptional point (pink line) to exactly at the exceptional point (dashed red line) to above the exceptional point (blue and cyan lines). Panel (b): the time $t_{\mathcal{P}}$, corresponding to the maximum of the instantaneous power $\mathcal{P}$ of the battery, as a function of the coupling-to-dissipation ratio $g/\gamma$ is represented with the pink-cyan line [cf. Eq.~\eqref{eq:adssdcdcads}]. Dashed lines: asymptotic results at small and large couplings [cf. Eq.~\eqref{eq:sfdfdxvxvsvefg} and Eq.~\eqref{eq:srvxvdxdvxdvgrs}]. Panel (c): the maximum instantaneous power $\mathcal{P} (t_{\mathcal{P}})$ as a function of $g/\gamma$ is represented with the pink-cyan line. Dashed lines: asymptotic results at small and large couplings [cf. Eq.~\eqref{eq:zdczdcsfddfs} and Eq.~\eqref{eq:sdfdzdczdcfsdf}]. Vertical grey lines in panels (b) and (c): guides for the eye at the exceptional point of $g = \gamma/4$ [cf. Eq.~\eqref{eq:sdfdfds}]. }
 \label{Pfig}
\end{figure*}

\noindent \textbf{Stored energy.} The energy $E$ stored in the battery holder follows from Eq.~\eqref{eq:xcxc} and the solutions of Eq.~\eqref{eq:zcszsc}, due to the conjugation relation $\langle b \rangle^\ast = \langle b^\dagger \rangle$. In the dissipationless limit, this stored energy follows the sinusoidal squared formula
\begin{equation}
\label{eq:sdfsdfdsfd}
\lim_{\gamma \to 0}E = \omega_b \: \Omega^2  \sin^2 \left( g t \right),  
\end{equation}
which highlights the intuitive properties of greater amounts of stored energy with stronger pulse strengths $\Omega$, and the reaching of an energetic maximum at some finite time (in this case, a maximal energy of $\omega_b \Omega^2$ is achieved after a time delay of $\pi/2g$ from the action of the pulse). However, in any realistic quantum battery the battery charger will suffer from nonzero dissipation $\gamma$, such that Eq.~\eqref{eq:sdfsdfdsfd} is superseded by
\begin{equation}
\label{eq:xcvxcv}
E = \begin{cases}
\omega_b \: \Omega^2 \left( \frac{g}{\Gamma} \right)^2 \sinh^2 \left( \Gamma t \right) \mathrm{e}^{- \frac{\gamma t}{2}}, & g < g_{\mathrm{EP}}, \\
  \omega_b \:  \Omega^2 \left(  \frac{\gamma t}{4} \right)^2 \mathrm{e}^{- \frac{\gamma t}{2}}, & g = g_{\mathrm{EP}}, \\
\omega_b \: \Omega^2  \left( \frac{g}{G} \right)^2 \sin^2 \left( G t \right) \mathrm{e}^{- \frac{\gamma t}{2}},  & g > g_{\mathrm{EP}},
\end{cases}
\end{equation}
with $G$ and $\Gamma$ as defined by Eq.~\eqref{eq:zxczxczxzx} and Eq.~\eqref{eq:opj} respectively, and where the exceptional point of Eq.~\eqref{eq:sdfdfds} separates the responses. Notably, the stored energy $E$ now decays with a time constant of $2/\gamma$, such that the battery charger should be disconnected from the battery holder after some finite amount of charging time to avoid complete degradation. Whilst connected, the battery holder dynamical energy $E$ displays a non-oscillatory (pink line) to oscillatory (blue and cyan lines) transition upon passing through its exceptional point (dashed red line), as is demonstrated graphically in Fig.~\ref{Efig}~(a) [cf. Eq.~\eqref{eq:xcvxcv}]. The maximum amount of stored energy in the battery, over all charging time $t$, allows for the optimal energetic charging time $t_{E}$ to be found, which satisfies $\mathrm{max}_t \{ E (t) \} = E (t_{E})$. The stationary points of Eq.~\eqref{eq:xcvxcv} imply the energetic times
\begin{equation}
\label{eq:xcdvzdvdvdvxcv}
t_E = \begin{cases}
\tfrac{\arctanh \left( \tfrac{4 \Gamma}{\gamma} \right)}{\Gamma}, & g < g_{\mathrm{EP}}, \\
\tfrac{4}{\gamma}, & g = g_{\mathrm{EP}}, \\
\tfrac{\arctan \left( \tfrac{4 G}{\gamma} \right)}{G},  & g > g_{\mathrm{EP}},
\end{cases}
\end{equation}
which is plotted as the pink-cyan line in Fig.~\ref{Efig}~(b), as a function of the coupling-to-dissipation ratio $g/\gamma$. Clearly, shorter optimal charging times $t_E$ arise when the charger-holder coupling $g$ is stronger, as was already implied in Fig.~\ref{Efig}~(a). In the two limiting cases of very weak ($g \ll g_{\mathrm{EP}}$) and very strong ($g \gg g_{\mathrm{EP}}$) coupling, Eq.~\eqref{eq:xcdvzdvdvdvxcv} reduces to its asymptotic forms
\begin{empheq}[left={t_E =\empheqlbrace}]{align}
   & \tfrac{4}{\gamma} \ln \left( \tfrac{\gamma}{2 g} \right), && g \ll g_{\mathrm{EP}}, \label{eq:srvgrs} \\
   & \tfrac{\pi}{2 g} - \tfrac{\gamma}{4g^2}, && g \gg g_{\mathrm{EP}}. \label{eq:sfdfdsvefg} 
\end{empheq}
The latter expression of Eq.~\eqref{eq:sfdfdsvefg} eventually recovers the inverse-$g$ behaviour of the fully dissipationless result of Eq.~\eqref{eq:sdfsdfdsfd}, and it is marked with a dashed blue line in Fig.~\ref{Efig}~(b). The former expression of Eq.~\eqref{eq:srvgrs} exudes a logarithmic character, as is distinguished by the dashed red line in Fig.~\ref{Efig}~(b), which suggests that the optimal charging time of very weakly coupled batteries is weakly (logarithmically) divergent. The exact quantity of energy $E (t_{E})$ stored in the quantum battery at the optimal time $t_E$ follows immediately from Eq.~\eqref{eq:xcvxcv} and Eq.~\eqref{eq:xcdvzdvdvdvxcv}. Explicitly, we find the brief formulae
\begin{equation}
\label{eq:xcxdxvdvxdvxdvvxcv}
E (t_{E}) = \begin{cases}
\omega_b \: \Omega^2 \: \mathrm{e}^{- \frac{\gamma}{2 \Gamma} \arctanh \left( \tfrac{4 \Gamma}{\gamma} \right) }, & g < g_{\mathrm{EP}}, \\
  \omega_b \: \Omega^2 \: \mathrm{e}^{-2}, & g = g_{\mathrm{EP}}, \\
\omega_b \: \Omega^2 \: \mathrm{e}^{- \frac{\gamma}{2 G} \arctan \left( \tfrac{4 G}{\gamma} \right) },  & g > g_{\mathrm{EP}},
\end{cases}
\end{equation}
where the number $\mathrm{e}^{-2} \simeq 0.135$ arises. This energy $E (t_{E})$ is graphed by the pink-cyan line in Fig.~\ref{Efig}~(c) as a function of the coupling-to-dissipation ratio $g/\gamma$. Notably, this optimal energy is bounded by zero and $\omega_b \Omega^2$ depending upon the size of the coupling strength $g$. The asymptotics of this optimal energy are provided by the simple expressions
\begin{empheq}[left={E (t_{E}) =\empheqlbrace}]{align}
   & \omega_b \: \Omega^2 \left( \tfrac{2 g}{\gamma} \right)^2, && g \ll g_{\mathrm{EP}}, \label{eq:sfddfs} \\
   & \omega_b \: \Omega^2 \left( 1  - \tfrac{\pi \gamma}{4 g} \right), && g \gg g_{\mathrm{EP}}. \label{eq:sdfdfsdf}
\end{empheq}
as drawn with dashed red and dashed blue lines respectively in Fig.~\ref{Efig}~(c). These formulae showcase the quadratic-in--$g$ decrease in optimal energy with very weak coupling, as well as the inverse-$g$ scaling with very strong coupling towards the upper energetic bound of $\omega_b \Omega^2$.
\\


\begin{figure*}[tb]
 \includegraphics[width=\linewidth]{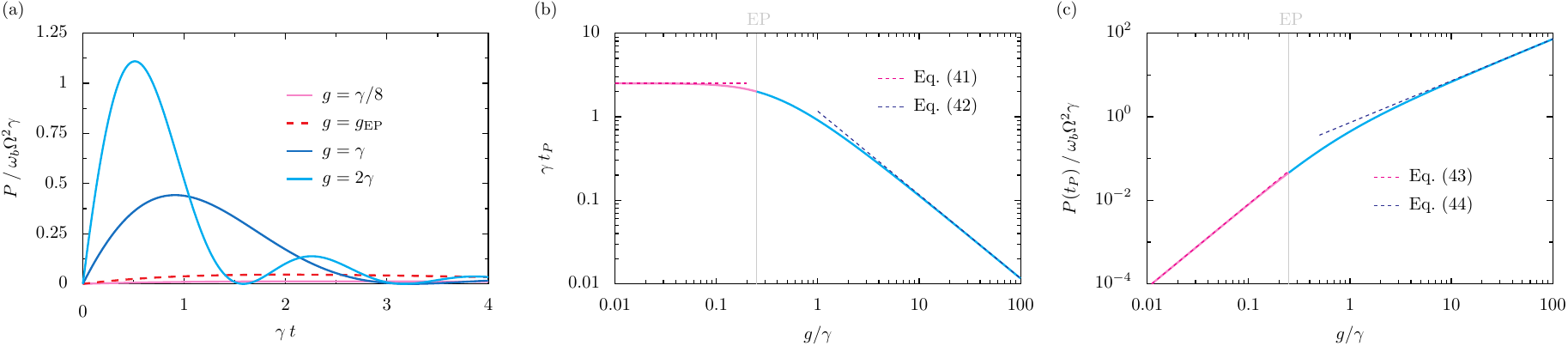}
 \caption{ \textbf{Average power of the pulsed quantum battery.} Panel (a): the average power $P$ of the battery (in units of $\omega_b \Omega^2 \gamma$) as a function of the time $t$ (in units of the inverse decay rate of the charger $\gamma^{-1}$) [cf. Eq.~\eqref{eq:xcxc3} with Eq.~\eqref{eq:xcvxcv}]. Coloured lines: results for increasingly large strengths of the charger-battery coupling $g$, from below the exceptional point (pink line) to exactly at the exceptional point (dashed red line) to above the exceptional point (blue and cyan lines). Panel (b): the time $t_{P}$, corresponding to the maximum of the average power $P$ of the battery, as a function of the coupling-to-dissipation ratio $g/\gamma$ is represented with the pink-cyan line [cf. Eq.~\eqref{eq:sdfsdfdf} and Eq.~\eqref{eq:sdfsddsvdfdf}]. Dashed lines: asymptotic results at small and large couplings [cf. Eq.~\eqref{eq:gjhgjghj} and Eq.~\eqref{eq:ghjhjg}]. Panel (c): the maximum average power $P (t_{P})$ as a function of $g/\gamma$ is represented with the pink-cyan line [cf. Eq.~\eqref{eq:xcxc3} with Eq.~\eqref{eq:sdfsdfdf} and Eq.~\eqref{eq:sdfsddsvdfdf}]. Dashed lines: asymptotic results at small and large couplings [cf. Eq.~\eqref{eq:zdczdscdcscsfddfs} and Eq.~\eqref{eq:sdccdsdcsdc}]. Vertical grey lines in panels (b) and (c): guides for the eye at the exceptional point of $g = \gamma/4$ [cf. Eq.~\eqref{eq:sdfdfds}]. }
 \label{PPfig}
\end{figure*}

\noindent \textbf{Instantaneous power.} The instantaneous power $\mathcal{P}$ of the pulsed quantum battery is obtained from Eq.~\eqref{eq:xcxc2} and the energetic results provided previously. In the idealized limit of zero dissipation, this instantaneous power is described by the simple sinusoidal formula  [cf. Eq.~\eqref{eq:sdfsdfdsfd}]
\begin{equation}
\label{eq:czxvcdscdcsdc}
\lim_{\gamma \to 0} \mathcal{P} =
\omega_b \: \Omega^2 g \sin \left( 2 g t \right).  
\end{equation}
This expression demonstrates the existence of an upper power bound of $\omega_b \Omega^2 g$, which is first reached at a time $\pi/4g$ after the occurrence of the driving pulse. The full driven-dissipative system has an instantaneous power instead encapsulated by [cf. Eq.~\eqref{eq:xcvxcv}]
\begin{widetext}
\begin{equation}
\label{eq:xcvzxczcxzxcv}
\mathcal{P} = \begin{cases}
\omega_b \: \Omega^2  \left( \frac{g}{\Gamma} \right)^2 \left[ \Gamma \sinh \left( 2 \Gamma t \right) - \frac{\gamma}{2} \sinh^2 \left( \Gamma t \right) \right] \mathrm{e}^{- \frac{\gamma t}{2}}, & g < g_{\mathrm{EP}}, \\
  \omega_b \: \Omega^2 \frac{\gamma}{2} \left( \frac{\gamma t}{4} \right) \left[ 1 - \frac{\gamma t}{4} \right] \mathrm{e}^{- \frac{\gamma t}{2}}, & g = g_{\mathrm{EP}}, \\
\omega_b \: \Omega^2  \left( \frac{g}{G} \right)^2 \left[ G \sin \left( 2 G t \right) - \frac{\gamma}{2} \sin^2 \left( G t \right) \right] \mathrm{e}^{- \frac{\gamma t}{2}},  & g > g_{\mathrm{EP}},
\end{cases}
\end{equation}
\end{widetext}
which features the common time constant $2/\gamma$, which controls the exponential decay. The instantaneous power $\mathcal{P}$ is plotted as function of time in Fig.~\ref{Pfig}~(a), presenting the critical behaviour of the model due to the distinct behaviour below (pink line), at (dashed red line) and above (cyan and blue lines) the exceptional point of Eq.~\eqref{eq:sdfdfds}. The trend of larger instantaneous power $\mathcal{P}$ with increasing large charger-holder coupling $g$, which peak at ever shorter timescales, is most apparent from the plot. The maximum of the instantaneous power $\mathcal{P}$ over all time $t$ suggests the optimal charging time $t_\mathcal{P}$, which is consistent with $\mathrm{max}_t \{ \mathcal{P} (t) \} = \mathcal{P} (t_{\mathcal{P}})$. Finding the relevant stationary points of Eq.~\eqref{eq:xcvzxczcxzxcv} leads to the optimal instantaneous power times $t_\mathcal{P}$, where
\begin{equation}
\label{eq:adssdcdcads}
t_\mathcal{P} = \begin{cases}
\tfrac{\arctanh \left( \tfrac{8 \Gamma \gamma - \sqrt{ \gamma^4- 256 g^4}  }{3 \gamma^2-16 g^2} \right)}{\Gamma}, & g < g_{\mathrm{EP}}, \\
  \tfrac{2 \left( 2-\sqrt{2} \right)}{\gamma}, & g = g_{\mathrm{EP}}, \\
\tfrac{\arctan \left( \tfrac{ \sqrt{256 g^4 - \gamma^4} - 8 G \gamma }{16 g^2 -3 \gamma^2} \right)}{G},  & g > g_{\mathrm{EP}},
\end{cases}
\end{equation}
where the number $2 \left( 2-\sqrt{2} \right) \simeq 1.17$ appears. The charging times $t_\mathcal{P}$ of Eq.~\eqref{eq:adssdcdcads} are given as the pink-cyan line in Fig.~\ref{Pfig}~(b) as a function of the coupling-to-dissipation ratio $g/\gamma$. Notably, the upper bound of $\ln ( 4 )/\gamma \simeq 1.39/\gamma$ is met with vanishing coupling $g$, while for very strong coupling $g$ the optimal time $t_\mathcal{P}$ is vanishing, with an inverse-$g$ relationship [cf. Eq.~\eqref{eq:czxvcdscdcsdc}]. These limiting cases are well-described by the asymptotics
\begin{empheq}[left={t_\mathcal{P} =\empheqlbrace}]{align}
   & \tfrac{\ln \left( 4 \right)}{\gamma} - 16 \left[ 1 - \ln \left( 2 \right) \right] \tfrac{g^2}{\gamma^3}, && g \ll g_{\mathrm{EP}}, \label{eq:sfdfdxvxvsvefg} \\
   & \tfrac{1}{4 g} \left( \pi - \tfrac{\gamma}{g} \right), && g \gg g_{\mathrm{EP}}. \label{eq:srvxvdxdvxdvgrs} 
\end{empheq}
which are plotted by the dashed red and dashed blue lines respectively in Fig.~\ref{Pfig}~(b), which exposes these key functional relationships. The progression of $\mathcal{P} ( t_\mathcal{P} )$, the instantaneous power at the optimal time $t_\mathcal{P}$ for the pulsed quantum battery, with increasing $g/\gamma$ is likewise plotted in Fig.~\ref{Pfig}~(c) [using Eq.~\eqref{eq:xcvzxczcxzxcv} with Eq.~\eqref{eq:adssdcdcads}]. The limiting cases are described by the compact expressions
\begin{empheq}[left={\mathcal{P} (t_{\mathcal{P}}) =\empheqlbrace}]{align}
   & \omega_b \: \Omega^2 \tfrac{g^2}{\gamma}, && g \ll g_{\mathrm{EP}}, \label{eq:zdczdcsfddfs} \\
   & \omega_b \: \Omega^2 g \left( 1 - \tfrac{\pi+2}{8} \tfrac{\gamma}{g} \right), && g \gg g_{\mathrm{EP}}. \label{eq:sdfdzdczdcfsdf} 
\end{empheq}
as denoted by the dashed red line and dashed blue line in Fig.~\ref{Pfig}~(c). In particular, the very weak coupling result and lower bound vanishes quadratically with $g$ [cf. Eq.~\eqref{eq:zdczdcsfddfs}], while the very strong coupling bound of $\omega_b \Omega^2 g$ is arrived at inverse-linearly in $g$ [cf. Eq.~\eqref{eq:sdfdzdczdcfsdf}].
\\


\noindent \textbf{Average power.} The average power $P$ of the pulsed quantum battery is defined by Eq.~\eqref{eq:xcxc3}, and it is made explicit by the energetic expressions given in Eq.~\eqref{eq:xcvxcv}. Upon completely neglecting dissipation from the battery charger ($\gamma \to 0$), the average power $P$ is simply given by [cf. Eq.~\eqref{eq:sdfsdfdsfd}]
\begin{equation}
\label{eq:czxfdvcdscdcsdc}
\lim_{\gamma \to 0} P =
\omega_b \: \Omega^2 \frac{ \sin^2 \left( g t \right)}{t},  
\end{equation}
which is indeed zero at $t = 0$ and later presents exact nodes at regular time intervals of $\pi/g$. The dynamic behaviour of Eq.~\eqref{eq:czxfdvcdscdcsdc} exhibits a rise to a global maxima at some time $t_P$, before gradually decaying inverse-linearly with time while oscillating. Working in the dimensionless variable $z = g t_P$, this maximum of average power $P(t_{P})$ can be found from the transcendental equation $\tan ( z) = 2 z$, where the range $0 < z < \pi/2$ emerges because the first maxima has the largest peak. Calculating a series expansion about the point $z = \pi/2$ up to 1st order and solving the resulting quadratic equation yields the approximate root $z = (2 \pi + \sqrt{ 9\pi^2 - 60})/10 \simeq 1.17$. Therefore, this idealized dissipationless case presents a maximum in average power at the time $t_P \simeq 1.17/g$, when the average power is $P (t_{P}) \simeq 0.72 \omega_b \Omega^2 g$. We plot the average powers $P$ as a function of time for the full driven-dissipative cases in Fig.~\ref{PPfig}~(a), using Eq.~\eqref{eq:xcxc3} with Eq.~\eqref{eq:xcvxcv}. As usual, the situations with the coupling $g$ below (pink line), at (dashed red line) and above (blue and cyan lines) the exceptional point at $g_{\mathrm{EP}}$ are graphed [cf. Eq.~\eqref{eq:sdfdfds}]. Intuitively, larger average powers $P$ arise with stronger couplings $g$, which are peaked at increasingly short periods of time. Above the exceptional point ($g > g_{\mathrm{EP}}$), nodes regularly occur at integer multiples of $\pi/G$ since no energy is stored in these special cases due to the Rabi-like oscillations [cf. Eq.~\eqref{eq:xcvxcv}]. The average power $P$ displays a global maximum at some particular time $t_P$, which is given by the solution of one of the following turning point equations (depending on the coupling strength $g$)
\begin{align}
\label{eq:sdfsdfdf}
  2 x &= \tanh \left( x\right) \left[ 1 + \tfrac{\gamma}{2\Gamma} x \right], &&g < g_{\mathrm{EP}}, \\
 2 y &= \tan \left( y\right) \left[ 1 + \tfrac{\gamma}{2G} y \right],  &&g > g_{\mathrm{EP}},  \label{eq:sdfsddsvdfdf}
\end{align}
which are written in the dimensionless variables $x = \Gamma t_P$ and $y = G t_P$, and where $0 < y < \pi/2$. The solutions of Eq.~\eqref{eq:sdfsdfdf} and Eq.~\eqref{eq:sdfsddsvdfdf} can be approximated by the analytical forms
\begin{equation}
\label{eq:zcszsczsczsc}
t_P \simeq \begin{cases}
\tfrac{\zeta - 8 \left( \zeta - 2 \right) \left(\frac{g}{\gamma} \right)^{3/2} }{\gamma}, & g < g_{\mathrm{EP}}, \\
 \tfrac{2}{\gamma}, & g = g_{\mathrm{EP}}, \\
\tfrac{Z - \frac{ 2 Z - 1 }{ 4 \sqrt{2} } \left(\frac{\gamma}{g} \right)^{3/4} }{g},  & g > g_{\mathrm{EP}},
\end{cases}
\end{equation}
where two transcendental numbers $\zeta$ and $Z$ have arisen, where
\begin{align}
      \zeta &= 2.512862...,  \label{eq:czxfdvsdfdfcdscdcsdc3} \\
  Z &= 1.165561....  \label{eq:czxfdvsdfdfcdscdcsdc} 
\end{align}
The exact optimal times $t_P$ are plotted as the pink-cyan line in Fig.~\ref{PPfig}~(b) as a function of the coupling-to-dissipation ratio $g/\gamma$, using the solutions of Eq.~\eqref{eq:sdfsdfdf} and Eq.~\eqref{eq:sdfsddsvdfdf}. In particular, the asymptotics of this important temporal quantity may be found from Eq.~\eqref{eq:zcszsczsczsc} as follows
\begin{empheq}[left={t_P = \empheqlbrace}]{align}
   & \tfrac{\zeta}{\gamma}, && g \ll g_{\mathrm{EP}}, \label{eq:gjhgjghj} \\
   & \tfrac{Z}{g}, && g \gg g_{\mathrm{EP}}, \label{eq:ghjhjg} 
\end{empheq}
which are plotted as the dashed red and dashed blue lines respectively in Fig.~\ref{PPfig}~(b). Notably, with very small coupling $g$ the optimal time $t_P$ tends to a constant (in units of $\gamma$), while for very large couplings $t_P$ drops off inverse-linearly with $g$. The maximum average power $P (t_P)$, as a function of the decisive ratio $g/\gamma$, is graphed in Fig.~\ref{PPfig}~(c) with the pink-cyan line. The scaling dependencies can be found explicitly through the limiting behaviours of Eq.~\eqref{eq:gjhgjghj} and Eq.~\eqref{eq:ghjhjg} as
\begin{empheq}[left={P (t_{P}) =\empheqlbrace}]{align}
   & \tfrac{ \sinh^2 \left( \frac{\zeta}{4} \right) \mathrm{e}^{-\frac{\zeta}{2} } }{\zeta } \: \omega_b  \Omega^2 \tfrac{\left( 4 g \right)^2 }{\gamma}, && g \ll g_{\mathrm{EP}}, \label{eq:zdczdscdcscsfddfs} \\
   & \tfrac{ \sin^2 \left( Z \right) }{Z} \: \omega_b \Omega^2 g, && g \gg g_{\mathrm{EP}}, \label{eq:sdccdsdcsdc}  
\end{empheq}
where the two numerical prefactors are $\sinh^2 ( \zeta /4  ) \exp (-\zeta/2) / \zeta \simeq 0.051$ and $\sin^2 ( Z )/Z \simeq 0.72$ respectively. Importantly, there is a quadratic scaling with $g$ at small coupling and a linear relationship with $g$ at large couplings. Taken together, Fig.~\ref{Efig}, Fig.~\ref{Pfig} and Fig.~\ref{PPfig} provide a comprehensive description of the energetics of the pulsed quantum battery, while brief discussions of the associated ergotropy and energy fluctuations are provided in the supplementary data.
\\


\noindent \textbf{Conclusion.} We have proposed a driven-dissipative quantum theory of a pulsed quantum battery. Our model features several desirable features, including brief analytical expressions for various energetic and temporal measures of the quality of the battery, which highlight the optimal setup of the system. We hope that our theory provides some utility for the expected deluge of experiments involving energy storage in quantum objects in the oncoming years~\cite{Quach2022, Joshi2022, Hu2022, Stevens2022, Maillette2022}, as well as stimulating further theoretical research into quantum technologies exploiting quantum continuous variables~\cite{Killoran2019, Fukui2022}.
\\


\noindent \textbf{Acknowledgments}\\
\textit{Funding}: CAD is supported by the Royal Society via a University Research Fellowship (URF\slash R1\slash 201158) and by Royal Society Enhanced Research Expenses which support MSU. CAD also gratefully acknowledges an Exeter-FAPESP SPRINT grant with the Universidade Federal de São Carlos. \textit{Discussions}: CAD thanks R.~Bachelard, A. Cidrim, A. C. Santos and C. J. Villas-Boas for fruitful discussions and for their generous hospitality during his visits to UFSCar.
\\


\begin{thebibliography}{100}




\bibitem{Alicki2013}
R.~Alicki and M.~Fannes,
Entanglement boost for extractable work from ensembles of quantum batteries,
\href{https://doi.org/10.1103/PhysRevE.87.042123}
{Phys. Rev. E \textbf{87}, 042123 (2013)}.

\bibitem{Campaioli2023}
For a review on quantum batteries, see:
F.~Campaioli, S.~Gherardini, J.~Q.~Quach, M.~Polini and G.~M.~Andolina,
Colloquium: quantum batteries,
\href{https://arxiv.org/abs/2308.02277}
{arXiv:2308.02277} [to appear in Reviews of Modern Physics].


\bibitem{Peterson2019}
J.~P.~S.~Peterson, T.~B.~Batalhão, M.~Herrera, A.~M.~Souza, R.~S.~Sarthour, I.~S.~Oliveira and R.~M.~Serra,
Experimental characterization of a spin quantum heat engine,
\href{https://doi.org/10.1103/PhysRevLett.123.240601}
{Phys. Rev. Lett. \textbf{123}, 240601 (2019)}.


\bibitem{Cimini2020}
V.~Cimini, S.~Gherardini, M.~Barbieri, I.~Gianani, M.~Sbroscia, L.~Buffoni, M.~Paternostro and F.~Caruso,
Experimental characterization of the energetics of quantum logic gates,
\href{https://doi.org/10.1038/s41534-020-00325-7}
{npj Quantum Inf. \textbf{6}, 96 (2020)}.

\bibitem{Quach2022}
J.~Q.~Quach, K.~E.~McGhee, L.~Ganzer, D.~M.~Rouse, B.~W.~Lovett, E.~M.~Gauger, J.~Keeling, G.~Cerullo, D.~G.~Lidzey and T.~Virgili,
Superabsorption in an organic microcavity: toward a quantum battery,
\href{https://doi.org/10.1126/sciadv.abk3160}
{Sci. Adv. \textbf{8}, eabk3160 (2022)}.

\bibitem{Joshi2022}
J~ Joshi and T.~S.~Mahesh,
Experimental investigation of a quantum battery using star-topology NMR spin systems,
\href{https://doi.org/10.1103/PhysRevA.106.042601}
{Phys. Rev. A \textbf{106}, 042601 (2022)}.


\bibitem{Hu2022}
C.-K.~Hu, J.~Qiu, P.~J.~P.~Souza, J.~Yuan, Y.~Zhou, L.~Zhang, J.~Chu, X.~Pan, L.~Hu, J.~Li, Y.~Xu, Y.~Zhong, S.~Liu, F.~Yan, D.~Tan, R.~Bachelard, C.~J.~Villas-Boas, A.~C.~Santos and D.~Yu,
Optimal charging of a superconducting quantum battery,
\href{https://doi.org/10.1088/2058-9565/ac8444}
{Quantum Sci. Technol. \textbf{7}, 045018 (2022)}.


\bibitem{Stevens2022}
J.~Stevens, D.~Szombati, M.~Maffei, C.~Elouard, R.~Assouly, N.~Cottet, R.~Dassonneville, Q.~Ficheux, S.~Zeppetzauer, A.~Bienfait, A.~N.~Jordan, A.~Auffèves and B.~Huard,
Energetics of a single qubit gate,
\href{https://doi.org/10.1103/PhysRevLett.129.110601}
{Phys. Rev. Lett. \textbf{129}, 110601 (2022)}.

\bibitem{Maillette2022}
I.~Maillette~de~Buy~Wenniger, S.~E.~Thomas, M.~Maffei, S.~C.~Wein, M.~Pont, N.~Belabas, S.~Prasad, A.~Harouri, A.~Lemaître, I.~Sagnes, N.~Somaschi, A.~Auffèves and P.~Senellart,
Experimental analysis of energy transfers between a quantum emitter and light fields,
\href{https://doi.org/10.1103/PhysRevLett.131.260401}
{Phys. Rev. Lett. \textbf{131}, 260401 (2023)}.



\bibitem{Ferraro2018}
D.~Ferraro, M.~Campisi, G.~M.~Andolina, V.~Pellegrini and M.~Polini,
High-power collective charging of a solid-state quantum battery,
\href{https://doi.org/10.1103/PhysRevLett.120.117702}
{Phys. Rev. Lett. \textbf{120}, 117702 (2018)}.


\bibitem{Andolina2018}
G.~M.~Andolina, D.~Farina, A.~Mari, V.~Pellegrini, V.~Giovannetti and M.~Polini,
Charger-mediated energy transfer in exactly solvable models for quantum batteries,
\href{https://doi.org/10.1103/PhysRevB.98.205423}
{Phys. Rev. B \textbf{98}, 205423 (2018)}.



\bibitem{Farina2019}
D.~Farina, G.~M.~Andolina, A.~Mari, M.~Polini and V.~Giovannetti,
Charger-mediated energy transfer for quantum batteries: an open-system approach,
\href{https://doi.org/10.1103/PhysRevB.99.035421}
{Phys. Rev. B \textbf{99}, 035421 (2019)}.




\bibitem{Crescente2020}
A.~Crescente, M.~Carrega, M.~Sassetti and D.~Ferraro,
Ultrafast charging in a two-photon Dicke quantum battery,
\href{https://doi.org/10.1103/PhysRevB.102.245407}
{Phys. Rev. B \textbf{102}, 245407 (2020)}.

\bibitem{NewCrescente2020}
A.~Crescente, M.~Carrega, M.~Sassetti and D.~Ferraro,
Charging and energy fluctuations of a driven quantum battery,
\href{https://doi.org/10.1088/1367-2630/ab91fc}
{New J. Phys. \textbf{22}, 063057 (2020)}.

\bibitem{Santos2020}
A.~C.~Santos, A.~Saguia and M.~S.~Sarandy,
Stable and charge-switchable quantum batteries,
\href{https://doi.org/10.1103/PhysRevE.101.062114}
{Phys. Rev. E \textbf{101}, 062114 (2020)}.


\bibitem{Carrega2020}
M.~Carrega, A.~Crescente, D.~Ferraro and M.~Sassetti,
Dissipative dynamics of an open quantum battery,
\href{https://doi.org/10.1088/1367-2630/abaa01}
{New J. Phys. \textbf{22}, 083085 (2020)}.

\bibitem{Santos2021}
A.~C.~Santos,
Quantum advantage of two-level batteries in the self-discharging process,
\href{https://doi.org/10.1103/PhysRevE.103.042118}
{Phys. Rev. E \textbf{103}, 042118 (2021)}.



\bibitem{Shaghaghi2002}
V.~Shaghaghi, V.~Singh, G.~Benenti and D.~Rosa,
Micromasers as quantum batteries,
\href{https://doi.org/10.1088/2058-9565/ac8829}
{Quantum Sci. Technol. \textbf{7}, 04LT01 (2022)}.

\bibitem{Gyhm2022}
J.-Y.~Gyhm, D.~Šafránek and D.~Rosa, 
Quantum charging advantage cannot be extensive without global operations,
\href{https://doi.org/10.1103/PhysRevLett.128.140501}
{Phys. Rev. Lett. \textbf{128}, 140501 (2022)}.



\bibitem{Catalano2023}
A.~G.~Catalano, S.~M.~Giampaolo, O.~Morsch, V.~Giovannetti and F.~Franchini,
Frustrating quantum batteries,
\href{https://arxiv.org/abs/2307.02529}
{arXiv:2307.02529}.



\bibitem{Friis2018}
N.~Friis and M.~Huber, 
Precision and work fluctuations in Gaussian battery charging,
\href{https://doi.org/10.22331/q-2018-04-23-61}
{Quantum \textbf{2}, 61 (2018)}.


\bibitem{Huangfu2021}
Y.~Huangfu and J.~Jing, 
High-capacity and high-power collective charging with spin chargers,
\href{https://doi.org/10.1103/PhysRevE.104.024129}
{Phys. Rev. E \textbf{104}, 024129 (2021)}.

\bibitem{Downing2023}
C.~A.~Downing and M.~S.~Ukhtary, 
A quantum battery with quadratic driving,
\href{https://doi.org/10.1038/s42005-023-01439-y}
{Commun. Phys. \textbf{6}, 322 (2023)}.

\bibitem{Centrone2021}
F.~Centrone, L.~Mancino and M.~Paternostro, 
Charging batteries with quantum squeezing,
\href{https://doi.org/10.1103/PhysRevA.108.052213}
{Phys. Rev. A \textbf{108}, 052213 (2023)}.




\bibitem{Konar2022}
T.~K.~Konar, A.~Patra, R.~Gupta, S.~Ghosh and A.~S.~De, 
Multimode advantage in continuous variable quantum battery,
\href{https://arxiv.org/abs/2210.16528}
{arXiv:2210.16528}.

\bibitem{Allahverdyan2004}
A.~E.~Allahverdyan, R.~Balian and T.~M.~Nieuwenhuizen,
Maximal work extraction from finite quantum systems,
\href{https://doi.org/10.1209/epl/i2004-10101-2}
{Europhys. Lett. \textbf{67}, 565 (2004)}.




\bibitem{Berry2003}
M.~V.~Berry,
Physics of nonhermitian degeneracies,
\href{https://doi.org/10.1023/B:CJOP.0000044002.05657.04}
{Czech. J. Phys. \textbf{54}, 1039 (2004)}.


\bibitem{Miri2019}
M.-A.~Miri and A.~Alu,
Exceptional points in optics and photonics,
\href{https://doi.org/10.1126/science.aar7709}
{Science \textbf{363}, 6422 (2019)}.

\bibitem{Downing2021}
C.~A.~Downing and V.~A.~Saroka,
Exceptional points in oligomer chains,
\href{https://doi.org/10.1038/s42005-021-00757-3}
{Commun. Phys. \textbf{4}, 254 (2021)}.

\bibitem{Bender2023}
C.~M.~Bender and D.~W.~Hook,
$\mathcal{PT}$-symmetric quantum mechanics,
\href{https://arxiv.org/abs/2312.17386}
{arXiv:2312.17386} [to appear in Reviews of Modern Physics].



\bibitem{Breuer2002}
H.-P.~Breuer and F.~Petruccione,
\textit{The Theory of Open Quantum Systems}
(Oxford University Press, Oxford, 2002).


\bibitem{Vidiella2023}
C.~A.~Downing and A.~Vidiella-Barranco,
Parametrically driving a quantum oscillator into exceptionality,
\href{https://doi.org/10.1038/s41598-023-37964-7}
{Sci. Rep. \textbf{13}, 11004 (2023)}.

\bibitem{Fox2023}
C.~A.~Downing and O.~I.~R.~Fox,
Unbalanced gain and loss in a quantum photonic system,
\href{https://doi.org/10.1088/2040-8986/ace5be}
{J. Opt. \textbf{25}, 095201 (2023)}.


\bibitem{Killoran2019}
N.~Killoran, T.~R.~Bromley, J.~M.~Arrazola, M.~Schuld, N.~Quesada and S.~Lloyd,
Continuous-variable quantum neural networks,
\href{https://doi.org/10.1103/PhysRevResearch.1.033063}
{Phys. Rev. Research \textbf{1}, 033063 (2019)}.


\bibitem{Fukui2022}
K.~Fukui and S.~Takeda,
Building a large-scale quantum computer with continuous-variable optical technologies,
\href{https://doi.org/10.1088/1361-6455/ac489c}
{J. Phys. B: At. Mol. Opt. Phys. \textbf{55}, 012001 (2022)}.





\end{thebibliography}
\end{document}